\documentclass[aps,pre,10pt,twocolumn,showpacs,superscriptaddress,preprintnumbers,amsmath,amssymb]{revtex4-1}

\usepackage{graphicx}
\usepackage{dcolumn}
\usepackage{bm}
\usepackage{times}

\begin{document}


\title{Glassy dynamics in relaxation of soft-mode turbulence}

\author{Fahrudin Nugroho}
\affiliation{Department of Applied Quantum Physics and Nuclear Engineering, Kyushu University, Fukuoka 819-0395, Japan}
\affiliation{Physics Department, Gadjah Mada University, Yogyakarta 55281, Indonesia}
\author{Takayuki Narumi}
\email{narumi@athena.ap.kyushu-u.ac.jp}
\affiliation{Department of Applied Quantum Physics and Nuclear Engineering, Kyushu University, Fukuoka 819-0395, Japan}
\author{Yoshiki Hidaka}
\affiliation{Department of Applied Quantum Physics and Nuclear Engineering, Kyushu University, Fukuoka 819-0395, Japan}
\author{Junichi Yoshitani}
\affiliation{Department of Applied Quantum Physics and Nuclear Engineering, Kyushu University, Fukuoka 819-0395, Japan}
\author{Masaru Suzuki}
\affiliation{Department of Applied Quantum Physics and Nuclear Engineering, Kyushu University, Fukuoka 819-0395, Japan}
\author{Shoichi Kai}
\affiliation{Department of Applied Quantum Physics and Nuclear Engineering, Kyushu University, Fukuoka 819-0395, Japan}

\date{\today}

\begin{abstract}
The autocorrelation function of pattern fluctuation is used to study soft-mode turbulence (SMT),
a spatiotemporal chaos observed in homeotropic nematics.
We show that relaxation near the electroconvection threshold deviates from the exponential.
To describe this relaxation, we propose a compressed exponential appearing in dynamics of glass forming liquids.
Our findings suggest that coherent motion contributes to SMT dynamics.
We also confirmed that characteristic time is inversely proportional to electroconvection's control parameter.
\end{abstract}


\maketitle

Spatiotemporal chaos has been observed and studied in spatially extended nonlinear systems \cite{Cross1993}.
More than a decade has passed since a spatiotemporal chaos was discovered
in homeotropic nematics \cite{Richter1995_EL, Richter1995_PRE, Kai1996, Hidaka1997_PRE}.
This type of spatiotemporal chaos was named soft-mode turbulence (SMT)
in reference to the softening of the state's macroscopic fluctuations \cite{Kai1996, Hidaka1997_PRE}.

SMT is a well-studied nonlinear phenomenon and remains of interest to many researchers.
Several studies have focused on the SMT relaxation characterized by the autocorrelation function 
\cite{Kai1996, Hidaka1997_PRE, Hidaka1997_JPSJ, Huh1998, Toth1998, Nagaya2000};
in these studies, the simple exponential has been conventionally employed to describe the relaxation.
Further, a softening relationship $\tau_{\text{s}} \propto \varepsilon^{-1}$ has been tested 
by using the time scale $\tau_{\text{s}}$ extracted from the simple exponential,
where $\varepsilon$ denotes the control parameter of the electroconvection.
However, deviation from the simple exponential has been suggested theoretically \cite{Tribelsky1999}.

We cautiously investigated the temporal autocorrelation function of SMT's pattern fluctuation,
and show that the relaxation deviates from the simple exponential decay
as the system approaches the electroconvection transition point ($\varepsilon\to+0$).
We thus find an alternative fitting form relating to the dynamics of glass forming liquid (GFL) 
and propose that there exists a similarity between SMT and GFL dynamics. 
GFL is an example of systems in which relaxation can be nonlinear \cite{Ediger2000, Debenedetti2001,Bouchaud2008}.
It is known that dynamical heterogeneity,
in which the spatiotemporal fluctuation due to the cooperative effects arises in local transient domains,
appears in the vicinity of the glass transition
and is a significant concept to understand mechanism of the glass transition \cite{Ediger2000, Weeks2000, Berthier2011}.
Several works have already pointed out similarity between chaotic and glassy dynamics qualitatively 
\cite{Tamura2002, Robledo2004_PhysLettA, Robledo2004_PhysicaA, Robledo2005, Baldovin2006};
our study experimentally provides quantitative evidences for the similarity.
SMT consists of randomly oriented roll domains
and shows patch structures \cite{Hidaka2006,Tamura2006} 
which may be strongly related to GFL properties such as the dynamical heterogeneity.
In this Letter, we discuss our results from these viewpoints.
Moreover, the softening relationship is verified with the time scale defined in the alternative equation.

In the systems in which the director ${\bf n}$ of a nematic liquid crystal is anchored perpendicular to the electrodes, 
continuous rotational symmetry exists on the plane ($x$--$y$ plane) parallel to the electrodes.
Applying a magnitude $V$ of the AC voltage above the Fr\'eedericksz threshold $V_{\text{F}}$, 
the director ${\bf n}$ tilts with respect to the $z$-axis
because the nematic liquid crystal we used has negative dielectric anisotropy,
thus breaking the continuous rotational symmetry. 
The projection ${\bf C}(\bm{r})$ of ${\bf n}$ onto the $x$--$y$ plane can freely rotate on the $x$--$y$ plane
because the azimuthal angle of ${\bf C}(\bm{r})$ behaves as Nambu--Goldstone mode \cite{Hertrich1992, Richter1995_EL, Tribelsky1996},
where $\bm{r}$ denotes the two-dimensional position vector on the $x$--$y$ plane.
Electroconvection occurs when the applied voltage is above the threshold value $V_{\text{c}}$, 
triggering the Carr--Helfrich instability \cite{deGennes1995}.
It interacts with ${\bf C}(\bm{r})$~\cite{Hidaka2006}, yielding SMT \cite{Kai1996, Rossberg1997}.
In this chaotic state,
the convective rolls were kept locally 
and their wave vector ${\bf q}(\bm{r})$ was isotropic over the whole system (see figures in Ref.~\cite{Kai1996}).

The experimental setup is similar to that in Ref.~\cite{Huh1998}, see Fig.~3 therein.
We carried out our experiments by using a nematic liquid crystal $N$--(4--Methoxybenzylidene)--4--butylaniline (MBBA) 
sandwiched between two glass plates with the distance $52 \pm 1$~$\mu$m.
We used circular electrodes, an indium tin oxide (ITO) with diameter $12.9$~mm, that is laid on the glass plates.
Statistical data were measured at the stabilized temperature $30.00 \pm 0.05^{\circ}$C. 
The dielectric constant parallel to the director was $\epsilon_{||} = 6.25 \pm 0.1$, 
and the electric conductivity parallel to the director was $\sigma_{||}= (1.17 \pm 0.1)\times10^{-7}~\Omega^{-1}$m$^{-1}$. 
The CCD camera (QImaging Retiga 2000R-Sy) mounted on the microscope
and the software (QCAPTURE PRO v.5) were used to capture pattern images on the $x$--$y$ plane.
The resultant image size was $1.14 \times 1.14$ mm$^{2}$ (i.e., $1000 \times 1000$ pixels). 

An AC voltage $V_{\text{AC}}(t)=\sqrt{2}V\cos(2\pi f t)$ was applied to the sample.
Two types of SMT pattern arise in response to a specific frequency;
namely oblique rolls in $f<f_{\text{L}}$ and normal rolls in $f>f_{\text{L}}$~\cite{Kai1996, Hidaka1997_PRE},
where $f_{\text{L}}$ denotes the Lifshitz frequency.
We set a fixed frequency of $f = 100$ Hz for SMT pattern to be the oblique roll, 
where $f_{\text{L}}=650$ Hz in our sample cell.
The voltage $V$ was first set between $V_{\text{F}} = 3.40$ V and  $V_{\text{c}} = 7.82$ V,
and we subsequently waited $10$ min until ${\bf C}(\bm{r})$-director reached a homogenous state
\footnote{$10$ min is sufficiently long for relaxation as shown in this Letter.}.
After the homogeneous state was reached, 
we instantaneously increased $V$ beyond $V_{\text{c}}$ 
in order to obtain the desired value of the normalized control parameter $\varepsilon=\left(V/V_{\rm c}\right)^{2}-1$ of the electroconvection,
and we then waited $20$ min to avoid a transient state.
$N_{T}=600$ images were captured every $\Delta T=1$ sec.

We observed in the above tests that SMT patterns were more disordered 
and their dynamics quickened with increasing $\varepsilon$, 
as previously discussed in Ref.~\cite{Hidaka1997_PRE}.
To extract such statistical properties,
we calculated the autocorrelation function $\tilde{Q}(\bm{r},\tau)$ of the pattern fluctuation in the Eulerian picture;
\begin{equation}
	\tilde{Q}(\bm{r},\tau) = \left< \Delta I(\bm{r},t+\tau)\Delta I(\bm{r},t)\right>_{t}\left< \Delta I(\bm{r},t)^{2}\right>_{t}^{-1}
\end{equation}
with $\Delta I(\bm{r},\tau):=I(\bm{r},\tau) - \left<I(\bm{r},t)\right>_{t}$,
where $I(\bm{r},t)$ denotes the intensity of transmitted light on a position $\bm{r}$ at time $t$
and the brackets $\left<~\right>_{t}$ averaging over time.
Specifically, each local function $\tilde{Q}(\bm{r},\tau)$ at 100 spatial points $\bm{r}$ was first calculated  
by averaging over the observation time $T=N_{T}\Delta T$. 
The obtained set of $\tilde{Q}(\bm{r},\tau)$ was then averaged over the $100$ spatial points 
to smooth errors caused by finite observation time.
The $10\times10$ points were equally spaced in the $x$--$y$ plane.

\begin{figure}[!t]
\includegraphics[width=0.98\linewidth]{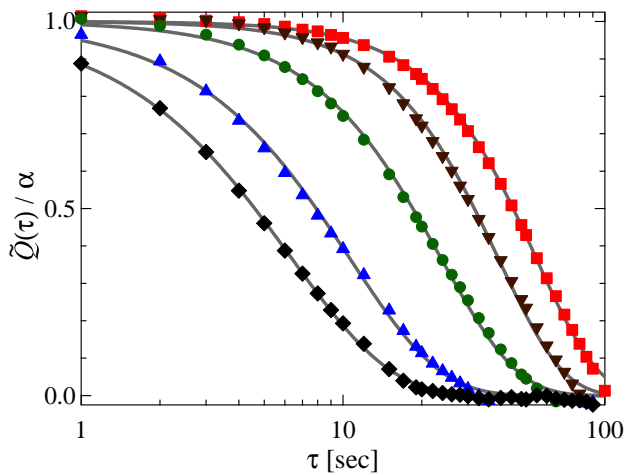}
\caption{(Color online) 
	Plot of the scaled autocorrelation function $\tilde{Q}(\tau) /\alpha$ 
	with the KWW fitting (gray line) for several values of the control parameters;
	$\varepsilon = 0.025$ (red square), $0.050$ (brown inverse triangle), $0.10$ (green circle), 
	$0.20$ (blue triangle), and $0.30$ (black diamond).
	Each normalization constant $\alpha$ was obtained by fitting.
}
\label{fig:TCor}
\end{figure}
{\it Result and Discussion.---}%
Experimental results of the temporal autocorrelation profiles are shown in Fig.~\ref{fig:TCor},
where $\tilde{Q}(\tau)$ denotes the spatially averaged $\tilde{Q}(\bm{r},\tau)$.
The sudden decrease appears in the short-time regime ($\tau < 1$~sec).
It is observed at low $\varepsilon$
and the mechanism has not been elucidated to date.
One may need measurements with smaller $\Delta T$ to clarify the specific actions of this mechanism;
we do not go into the details of this regime in this Letter.
Note that GFL also present such a short-time dynamics \cite{Ediger2000, Debenedetti2001}
and natures of the short-time dynamics do not affect 
universal behavior of the long-time regime \cite{Gleim1998, Tokuyama2007}.

Temporal autocorrelation functions have been conventionally described 
in terms of an exponential decay 
$\tilde{Q}(\tau) \propto \exp\left[-|\tau|/\tau_{\text{s}}\right]$.
If the dynamics of the pattern fluctuation is Markovian,
then the relaxation would be described by the simple exponential.
Here, we carefully analyzed the relaxation behavior
and found that the simple exponential is not suitable for describing $\tilde{Q}(\tau)$ at a low $\varepsilon$,
meaning that SMT dynamics at a low $\varepsilon$ is dominated by the non-Markov process.
It is difficult to identify the relaxation form {\it ab initio};
nevertheless, it may be possible to make predictions to some extent from experimental results.
First, one should take into account the fact
that SMT relaxations at a high $\varepsilon$ are well-described by the simple exponential.
In addition, there is no sign of any transition above the electroconvection transition.
Thus, it is expected 
that the SMT relaxation for whole $\varepsilon>0$ can be described by an individual function
that converges to the simple exponential at a high $\varepsilon$.
One possibility for such a fitting function is
the Kohlrausch--Williams--Watts (KWW) equation \cite{Kohlrausch1854, Williams1970}
\begin{equation}
	\tilde{Q}(\tau) =\alpha \exp\left[-\left(|\tau| \left/ \tau_{\text{KWW}} \right. \right) ^{\beta} \right],
	\label{KWW}
\end{equation}
where $\alpha$ denotes a normalized constant, $\beta$ the so-called Kohlrausch exponent,
and $\tau_{\text{KWW}}$ the relaxation time of the KWW function.
The coefficient $\alpha$ was introduced in Eq.~(\ref{KWW}) to eliminate the short-time dynamics.
The KWW relation is empirical for relaxation in disordered systems (e.g., Ref.~\cite{Bouchaud2008} as a review).
As illustrated in Fig.~\ref{fig:TCor},
the KWW relation well describes the SMT dynamics for whole $\varepsilon$
\footnote{We confirmed the appearance of the linear slope in log-log plot of $-\ln \tilde{Q}(\tau)$ versus $\tau$,
implying that the SMT autocorrelation function can be well-quantified by the KWW expression.}.

\begin{figure}[!t]
\includegraphics[width=0.85\linewidth]{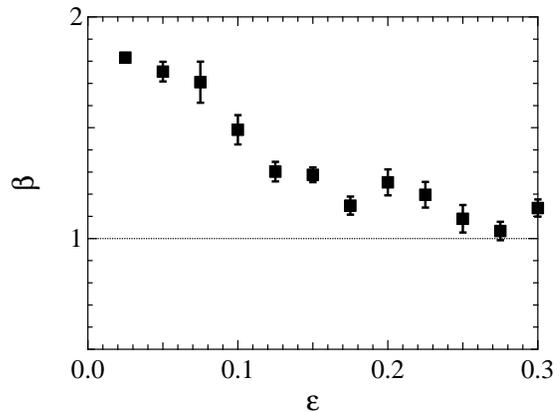}
\caption{
	An $\varepsilon$-dependence of $\beta$ in Eq.~(\ref{KWW}).
	The dotted line corresponds to the simple exponential.
	Standard deviation is used as error bar.
}
\label{fig:KWW_beta}
\end{figure}
Figure \ref{fig:KWW_beta} shows an $\varepsilon$-dependence of the exponent $\beta$.
The KWW equation is called the compressed exponential for $\beta > 1$ or the stretched exponential for $\beta < 1$.
In the investigated control parameter range, 
a deviation from simple exponential behavior (i.e., $\beta=1$) has been observed in a low $\varepsilon$.
The deviation seems to increase monotonically with decreasing $\varepsilon$;
no jumps nor convergence were found in our results at a low $\varepsilon$.
Note that the exponent $\beta\simeq 3/2$ has been suggested in a microscopic model \cite{Cipelletti2000, Bouchaud2001, Cipelletti2003},
where the dynamics is mainly dominated by the random appearance of localized rearrangements;
however, no specific features were observed at $\beta \simeq 3/2$ in our results.
On the other hand, it is reasonable to expect that 
the exponent $\beta$ converges to unity with increasing $\varepsilon$
because the temporal autocorrelation function can be fitted by the simple exponential at high control parameters,
i.e., the stretched exponential is not observed in SMT.
This convergence at a high $\varepsilon$ was also confirmed via power-spectrum analysis.

It remains unclear to which models the compressed exponential corresponds.
It has been, however, reported that the compressed exponential appears
as a common feature of jammed, non-diffusive systems 
\cite{Cipelletti2000, Bellour2003, Falus2006, Robert2006, Fluerasu2007, Caronna2008}.
In particular, Caronna et al.~have shown 
a deviation from the simple exponential as the system approaches the glass transition point \cite{Caronna2008}.
In addition, Bouchaud has stated 
that the compressed exponential deeply relates to the dynamical heterogeneity \cite{Bouchaud2008}.

Does the cooperative effect exist in SMT?
It has been also reported that, in SMT,
there exist patch domains 
in which the roll direction ${\bf q}(\bm{r})/|{\bf q}(\bm{r})|$ is unique \cite{Hidaka2006, Tamura2006}.
Since each patch domain is regarded as being stable,
the domain should keep its size,
and then the rolls in a patch should move cooperatively.
The ballistic-like, coherent movement of the pattern is thus likely to occur in SMT,
and the mechanical coherency is classified as a non-Markov process.
That is, the memory effects cannot be negligible.
According to the fluctuation-dissipation relation,
our result can also be interpreted as showing the emergence of non-thermal noise in SMT, 
as supported by the Lagrangian picture \cite{Hidaka2006, Hidaka2010}.
It is consequently noteworthy that 
the non-thermal fluctuation caused by the patch structure leads to non-exponential relaxation.
This argument is supported by the fact that the average patch size increases with decreasing $\varepsilon$ \cite{Tamura2006},
which is consistent with the results shown in Fig.~\ref{fig:KWW_beta}.
In dynamics of GFL, the length scale characterizing dynamical heterogeneity also increases 
as the system approaches the glass transition point \cite{Berthier2005}.
On the other hand, as $\varepsilon$ increases in SMT,
roll structures are easily disturbed 
and then the lifetime of a roll structure shortens.
The ballistic dynamics eventually eliminates at a high $\varepsilon$,
leading to the memory function represented by the delta function;
i.e., relaxation can be described by the simple exponential at a high $\varepsilon$.
It is also consistent that dynamical heterogeneity lifetime shortens 
as the system departs from the glass transition \cite{Mizuno2010}. 

%
\begin{figure}[!t]
\includegraphics[width=0.85\linewidth]{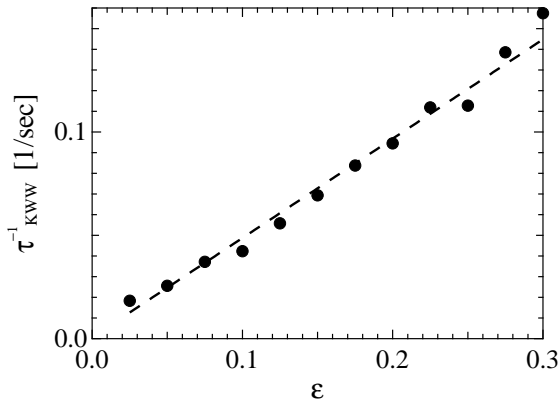}
\caption{
	An $\varepsilon$-dependence of $\tau_{\text{KWW}}$ in Eq.~(\ref{KWW}).
	The dashed line represents a fitting line: 
	$\tau_{\text{KWW}}^{-1} = c_{0}+c_{1}\varepsilon$
	with $c_{0} = 0.00 \pm 0.0031$ sec$^{-1}$ and $c_{1} = 0.48 \pm 0.017$ sec$^{-1}$.
}
\label{fig:KWW_tau}
\end{figure}
We next discuss the correlation time $\tau_{\text{KWW}}$.
Kai et al.~have experimentally shown that the correlation time $\tau_{\text{s}}$ diverges as $\tau_{\text{s}}\propto \varepsilon^{-1}$ \cite{Kai1996}.
It has been discussed that the relationship between correlation time and the control parameter 
corresponds to supercriticality of the SMT bifurcation;
this was employed as evidence that softening of the irregular modes occurs at $\varepsilon=0$. 
However, since we have revealed that the physically meaningful time scale is not $\tau_{\text{s}}$ but $\tau_{\text{KWW}}$,
we should reexamine whether the softening relation with $\tau_{\text{KWW}}$ holds.
The result obtained from the compressed exponential function satisfies 
the relationship $\tau_{\text{KWW}}\propto \varepsilon^{-1}$ shown in Fig.~\ref{fig:KWW_tau};
namely, $c_{0} = 0.00$ sec$^{-1}$ defined in $\tau_{\text{KWW}}^{-1} = c_{0}+c_{1}\varepsilon$.
It is therefore correct, 
even if the compressed exponential is used instead of the simple exponential, 
that the bifurcation to SMT occurs supercritically 
and that the bifurcation point coincides with electroconvection threshold $\varepsilon=0$.

{\it Concluding Remarks.---}%
In this study, we experimentally examined 
the temporal autocorrelation function of pattern fluctuation in SMT,
a spatiotemporal chaos in the homeotropic nematics.
Our research revealed that
the relaxation deviates from the simple exponential decay 
as the system approaches the transition point $V_{\text{c}}$ of the electroconvection from above,
implying that the pattern fluctuation is not perfectly randomized nor Markovian at a low $\varepsilon$.
We have proposed the compressed exponential (i.e., the KWW equation of $\beta \ge 1$)
as an alternative fitting form for whole $\varepsilon$.
Although the KWW function was introduced as heuristic fitting form,
it well describes the SMT relaxation as shown in Fig.~\ref{fig:TCor}.
Few models showing the compressed exponential currently exist in the literature.
Our finding can be thus regarded as a convenient observation of random motion 
as well as jammed, non-diffusive systems.

The Kohlrausch exponent $\beta$ deviates from unity as $\varepsilon$ approaches zero.
Similar deviation has been observed in GFL in the vicinity of the glass transition point.
Thus, we have discovered a possible method for relating SMT to GFL dynamics from the viewpoint of the coherent structure.
In SMT, the ballistic-like motion on the roll structure generates such a coherent motion,
and then non-exponential relaxation subsequently emerges from the memory effect.
This is just a possibility at the moment; further study is needed.
We presently investigate the memory function 
calculated from the correlation function \cite{Narumi_tobe_SMTMem}
to directly clarify the memory effect in SMT.
In addition, this may represent a possible way to measure SMT statics
by well-studied physical quantities in the study of the glass transition
such as dynamical correlation functions \cite{Donati1999, Weeks2007, Narumi_BCM_2010}
and the four-point correlation function (Ref.~\cite{Berthier2005} and references therein).

Through this research, 
we have also confirmed the $\varepsilon$-dependence of the correlation time $\tau_{\text{KWW}}$ extracted from the KWW function.
The time scale is inversely proportional to the control parameter and diverges at $\varepsilon\to +0$,
which corresponds to our expected outcome that SMT continuously (i.e., supercritically) arises at $\varepsilon=0$.

As a final remark, 
we provide a comment about theoretical works of the spatiotemporal chaos.
Several publications focusing on universality in the spatiotemporal chaos 
\cite{Tribelsky1996, Fujisaka2003, Tanaka2005, Tanaka2007, Tribelsky2008, Tanaka2010} 
have noted the similarity between SMT and the Nikolaevskii turbulence (NT),
a theoretical model of spatiotemporal chaos 
described by the so-called Nikolaevskii equation \cite{Nikolaevskii1989}.
The NT's spatiotemporal chaos is also generated supercritically as well as SMT.
Tanaka and Okamura \cite{Tanaka2010} have recently reported that,
in the one-dimensional NT that can be regarded as one-dimensional SMT, 
the temporal autocorrelation function is represented not by the simple exponential but by the power law.
We thus tried to fit our data by the power law;
however, the power law is not suitable for our two-dimensional SMT.
To clarify details,
one should investigate modal (i.e., wave-number dependent) autocorrelation functions.
The temporal autocorrelation function shown in this Letter is regarded 
as a wave-number average of the modal one \cite{Tanaka2010}.
We will discuss SMT's modal autocorrelation function elsewhere \cite{Narumi_tobe_SMTMem}.
In addition, spatial dimension might have an effect on the relaxation shape;
future research should tackle differences related to the spatial dimension.
Further advancement of higher-dimensional models 
such as those described in Refs.~\cite{Fujisaka2003,Tanaka2007} is similarly desired.

The authors acknowledge Prof.~Mikhail~I.~Tribelsky for productive discussions.
This work was partially supported by KAKENHI (Nos.~20111003, 21340110, and 21540391) 
and the JSPS Core-to-Core Program
"International research network for non-equilibrium dynamics of soft matter."
F.~N.~acknowledges the support of The Directorate General of Higher Education, 
Department of National Education of Indonesia. 

\bibliography{./SMT_TotalCorr_arXiv.bbl}

\end{document}